\documentclass[pdflatex,sn-basic]{sn-jnl}% Basic Springer Nature Reference Style/Chemistry Reference Style

\usepackage{graphicx}
\usepackage{subfigure}
\usepackage{amsthm}
\usepackage{mathrsfs}
\usepackage{amssymb}
\usepackage{bm} 

\usepackage{lineno}   %% <- no mathlines option
\usepackage{amsmath}  %% <- after lineno
\usepackage{etoolbox} %% <- for \cspreto, \csappto

\newcommand*\linenomathpatch[1]{%
  \cspreto{#1}{\linenomath}%
  \cspreto{#1*}{\linenomath}%
  \csappto{end#1}{\endlinenomath}%
  \csappto{end#1*}{\endlinenomath}%
}

\linenomathpatch{equation}
\linenomathpatch{gather}
\linenomathpatch{multline}
\linenomathpatch{align}
\linenomathpatch{alignat}
\linenomathpatch{flalign}

\jyear{2023}%

\theoremstyle{thmstyleone}%
\theoremstyle{thmstyletwo}%

\theoremstyle{thmstylethree}%

\begin{document}

% old title: Mechanisms of spike in the brain boost the robustness under attacks
% Mechanisms of spike boost robustness

%\title[Robustness of SNN against attacks]{Brain-like spiking neural networks allow robustness against attacks}

\title[Robustness of SNNs against attacks]{Spike timing reshapes robustness against attacks in spiking neural networks
%Timing is all you need for robustness against attacks in spiking neural networks
}


\author[1]{\fnm{Jianhao} \sur{Ding}}\email{djh01998@stu.pku.edu.cn}

\author*[1,2]{\fnm{Zhaofei} \sur{Yu}}\email{yuzf12@pku.edu.cn}

\author[1]{\fnm{Tiejun} \sur{Huang}}\email{tjhuang@pku.edu.cn}

\author[3]{\fnm{Jian K.} \sur{Liu}}\email{j.liu9@leeds.ac.uk}

\affil[1]{\orgdiv{School of Computer Science}, \orgname{Peking University}, \orgaddress{\postcode{100871}, \state{Beijing}, \country{China}}}

\affil[2]{\orgdiv{Institute for Artificial Intelligence}, \orgname{Peking University}, \orgaddress{\postcode{100871}, \state{Beijing}, \country{China}}}

\affil[3]{\orgdiv{School of Computing}, \orgname{University of Leeds}, \orgaddress{\postcode{LS2 9JT}, \state{Leeds}, \country{United Kingdom}}}


%\abstract{The success of deep learning is partially shrouded in the shadow of adversarial attacks. In contrast, the brain is far more robust at complex cognitive tasks. Some neurons in the brain communicate via electrical conduction which are similar to neurons in artificial neural networks. Most neurons in the nervous system rely on spikes to transmit information. However, it is not fully understood about whether or not the mechanism of spiking improve the robustness of the system. We compared the performance of spiking neural networks and analog neuron networks under attacks, and found that the spiking neural networks can achieve robustness improvement through neuronal encoding, neuronal decoding, and learning rules. In particular, the robustness improvement is fairly obvious when using temporal coding and refined training methods. We suggest that the utility of complex coding and synaptic learning schemes in the brain is not just a trade-off between information transmission and reaction speed, but also a consideration of reliability of complex perceptual functions.}

\abstract{The success of deep learning in the past decade is partially shrouded in the shadow of adversarial attacks. In contrast, the brain is far more robust at complex cognitive tasks. Utilizing the advantage that neurons in the brain communicate via spikes, spiking neural networks (SNNs) are emerging as a new type of neural network model, boosting the frontier of theoretical investigation and empirical application of artificial neural networks and deep learning. Neuroscience research proposes that the precise timing of neural spikes plays an important role in the information coding and sensory processing of the biological brain. However, the role of spike timing in SNNs is less considered and far from understood. Here we systematically explored the timing mechanism of spike coding in SNNs, focusing on the robustness of the system against various types of attacks. We found that SNNs can achieve higher robustness improvement using the coding principle of precise spike timing in neural encoding and decoding, facilitated by different learning rules. Our results suggest that the utility of spike timing coding in SNNs could improve the robustness against attacks, providing a new approach to reliable coding principles for developing next-generation brain-inspired deep learning. }




\maketitle

\section{Introduction}\label{sec1}

The question of how the brain processes information continues to be an intriguing topic of investigation within the field of neuroscience and artificial intelligence~\citep{Borst1999, Quiroga2009, Yamins2016, Zador2023}. Conventionally, it has been believed that neurons encode sensory stimuli by means of the number of spikes, known as firing rate coding, a concept that has been supported by studies dating back to the early 20th century~\citep{Adrian1926}. It has been identified in multiple sensory systems, such as the motor~\citep{srivastava2017motor} and visual cortex~\citep{berry1997structure, walker2020neural}, and has been used to explain slow sensory responses in the brain~\citep{lu2001temporal, prescott2008spike}. The term rate coding has a rich meaning covering population coding~\citep{auge2021survey}. However, recent research conducted in the past decade has suggested that spike timing also plays a pivotal role in the coding of not only sensory stimuli but also decision variables~\citep{Stein2005, Gollisch2008, Quiroga2009}. Consequently, a notion has arisen for a combination of neuronal coding for perception and decision-making that incorporates both rate and timing coding in a coherent fashion~\citep{mehta2002role, Hong2016, Lankarany2019}. 

In the realm of neural network models, artificial neural networks (ANNs) may be deemed a form of rate coding due to their simulation of neuronal systems~\citep{fukushima1975cognitron}. Spiking neural networks (SNNs), on the other hand, have emerged as brain-inspired neural network models that utilize spiking neural units to replace the neural model in ANNs~\citep{maas1997networks, bertens2020network, wozniak2020deep}. Despite this, most SNNs still utilize firing rates within a designated time frame, allowing for the utilization of established ANNs in computing~\citep{sengupta2019going, rueckauer2017conversion}. While certain SNN studies have attempted to use temporal spiking coding, they have been limited to the consideration of several spikes~\citep{park2020t2fsnn, zhou2021temporal, Stoeckl2021, goltz2021fast}, without computing firing rates. Therefore, the question of how to implement a hybrid neural coding scheme that incorporates both firing rate and spike timing in SNNs remains unclear, with the role of spike timing in SNNs still largely unexplored. In this study, we propose a novel neural coding approach that embeds the sequence of spiking timing into firing rate SNN models. This approach builds upon the current advancements in rate coding of SNN models while providing a new perspective on the role of spike timing beyond single spikes.

We demonstrate the efficacy of our proposed approach in a series of tasks that aim to enhance robustness against attacks. Our brains possess remarkable abilities to perform a diverse range of functions robustly, even in the face of ever-changing environments~\citep{deneve2017brain}. In contrast, current deep-learning models of ANNs lack robustness when faced with adversarial attacks. Even minor modifications to input images that are readily apparent to the human eye can cause ANNs to produce inaccurate predictions~\citep{goodfellow2014explaining}. Although recent advancements have allowed for the training of biologically interpretable SNNs with high precision~\citep{wu2019direct, yin2021accurate, fang2021deep}, the susceptibility of SNNs to adversarial attacks remains poorly understood~\citep{ding2022snn}. In this study, we investigated the potential benefits of the spiking timing mechanism as compared to ANNs, in terms of neural encoding and decoding, through the use of various types of learning rules. Our findings from these multiple perspectives suggest that the incorporation of spike timing mechanisms can significantly enhance the robustness of SNNs, particularly in the context of adversarial attacks on SNN learning.

\section{Results}\label{sec2}

\subsection{Rate-based SNNs gain robustness through synchronization schemes.}

\begin{figure}[t]
\begin{center}
\centerline{\includegraphics[width=1.0\columnwidth]{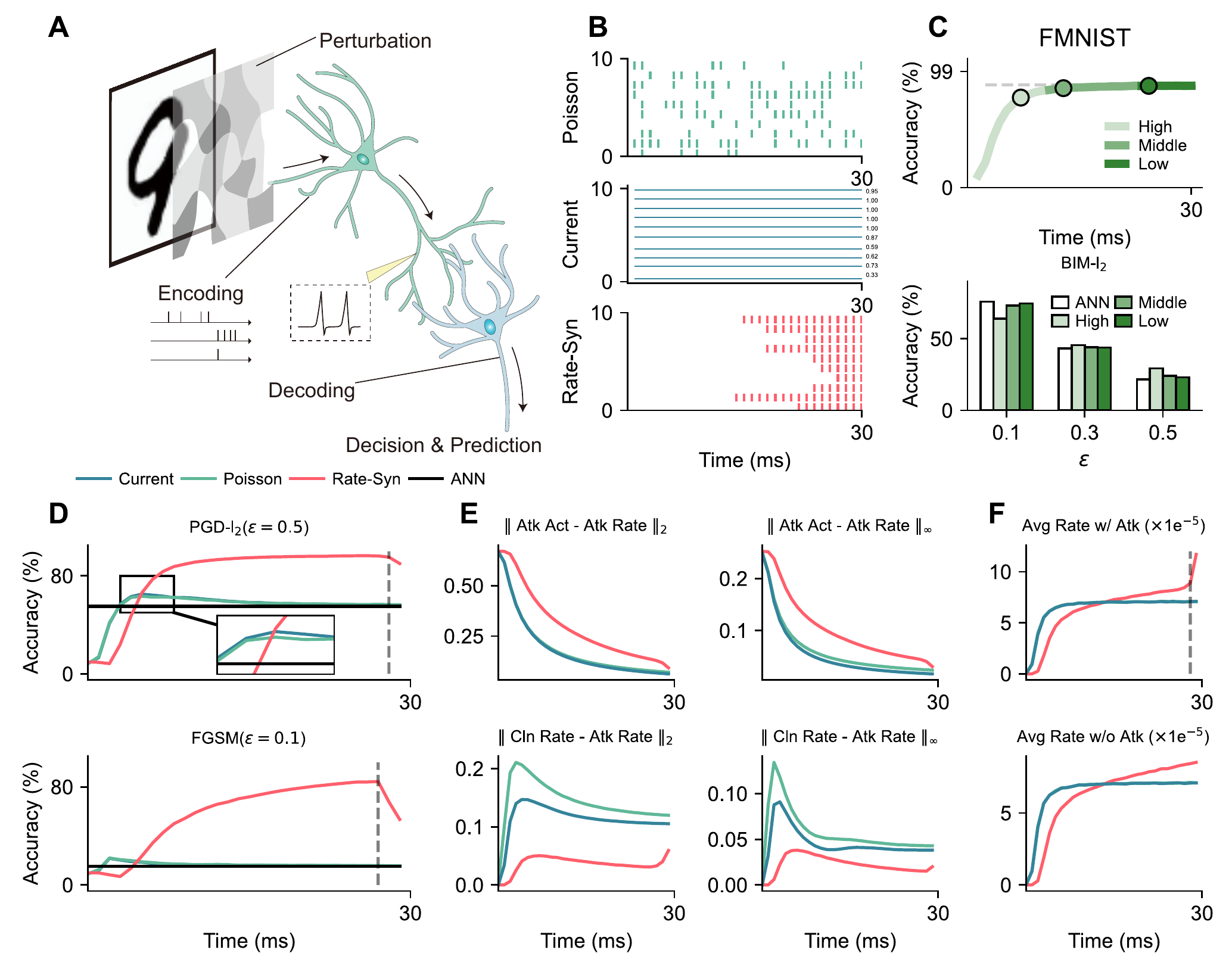}}
\caption{Illustration of the robustness of SNN with regard to signal encoding. (A) When encountering adversarial examples, biologically realistic SNNs can boost robustness by encoding spike patterns and decoding neuronal output to make predictions or decisions. (B) Illustration of three input encoding patterns: Current coding, Poisson coding, and Rate-Syn coding. (C) Accuracy of converted rate-based SNN with regard to simulation time and the corresponding accuracy under white-box BIM-$l_2$ attack. (D) Accuracy of three input encoding schemes under white-box PGD-$l_2$ and FGSM attacks with regard to simulation time. (E) Distance between the firing rate of the attacked SNN (Atk Rate) and the attacked ANN activation (Atk Act) of the hidden neurons, and the distance between the firing rate of the attacked SNN (Atk Rate) and the clean SNN (Cln Rate) of the hidden neurons. (F) Average instantaneous firing rate (Avg Rate) of the hidden neurons under or without attack.}
\label{fig:part1}
\end{center}
\end{figure}

Rate coding mainly uses spike frequency to encode information~\citep{Adrian1926}. 
%It has been identified in multiple sensory systems, such as the motor~\citep{srivastava2017motor} and visual cortex~\citep{berry1997structure, walker2020neural}, and has been used to explain slow sensory responses in the brain~\citep{lu2001temporal, prescott2008spike}. The term rate coding has a rich meaning covering population coding~\citep{auge2021survey}. 
Here, the meaning of `rate' term is limited to the average firing rate within a time window. Recent advancement in SNNs has shown that an ANN can be transferred to a rate-based SNN through weight rescaling~\citep{cao2015spiking, diehl2015fast}. Based on the conversion theory, SNN encode information mainly in the firing rate, which approximates the activation of ANNs simulating in limited discrete time steps. To verify the advantages against perturbation (or attack) brought by the discretization of spike timing, we trained two three-layered ANNs with 100 hidden neurons to perform classification tasks on the MNIST and Fashion MNIST (FMNIST) datasets and convert them to SNNs with integrate-and-fire (IF) neurons. This makes it fair to compare the differences between SNNs adopting variant rate coding schemes (Fig.~\ref{fig:part1}(B)) with extended time dimensions and ReLU-based ANNs. Besides, the conversion allows us to conduct white-box attacks from the vanilla ANN with differentiable activation.

% To verify the advantages on the perturbation brought by this discretization, we trained three-layered ANNs with 100 hidden neurons to perform classification tasks on the MNIST and Fashion MNIST datasets and converted them to SNNs with integrate-and-fire models (IF model). This makes it fair to compare the differences between SNNs adopting variant rate coding scheme (Fig.~\ref{fig:part1}(B)) with extended time dimensions and ReLU-based ANNs. Besides, the conversion give us opportunity to conduct white-box attacks from the vanilla ANN with differentiable activation.

We first considered Poisson coding, in which the converted SNNs receive Bernoulli-sampled spikes of fixed frequency proportional to the pixel intensity. Previous research has proved that the performance of the converted SNNs increases with the simulation time~\citep{rueckauer2017conversion}. Here we tested whether the robustness of the converted SNNs with Poisson coding is related to the simulation time. According to the time-varying conversion loss (ANN accuracy minus SNN accuracy), we manually selected three simulation time values to test the robustness of SNNs of ``high accuracy loss'' (High), ``low accuracy loss'' (Middle), and ``high accuracy'' (Low). For the Fashion MNIST dataset, the three values are 6 ms, 12 ms, and 24 ms~(Fig.~\ref{fig:part1}(C)). Then we performed white-box iterative gradient adversarial attacks on these SNNs. We observed a slight increase of robustness of SNN with ``high accuracy loss'' when the attack intensity $\varepsilon$ is large (=0.5). However, the overall robustness of Poisson coding is negligible. 

% The conversion theory implies that fewer timesteps for inferencing SNN will cause accuracy drop~\citep{rueckauer2017conversion}. Hence, we first examined the effect of timesteps. The converted SNNs received bernoulli-sampled pulses of 1kHz (also known as Poisson coding) from pixel intensity. According to the time-varing conversion loss (ANN accuracy minus SNN accuracy), we manually selected three simulation time values to test the robustness of SNNs of ``high accuracy loss'', ``low accuracy loss'', and ``high accuracy''. For Fashion MNIST dataset, the three values are 6, 12, 24~(Fig.~\ref{fig:part1}(C)). And we performed white-box iterative gradient adversarial attacks on these SNNs. We observed a slight increase of robustness of SNN with ``high accuracy loss'' when $\varepsilon$ is large (=0.5). The overall robustness for Poisson coding is negligible. 

Apart from Poisson coding, the converted SNN also allows direct input of constant ``current'',  denoted as `Current coding'. This coding strategy can bring better accuracy for converted SNNs~\citep{sharmin2020inherent, ding2021optimal} and thus becomes one of the most popular rate coding schemes. To compare Poisson coding and Current coding, we plotted the time-varying classification accuracy on the MNIST dataset under white-box PGD-$l_2$($\varepsilon=0.5$) and FGSM($\varepsilon=0.1$) attack (Fig.~\ref{fig:part1}(D)). The tendencies of the curves of SNN fed by Poisson and Current coding are almost the same. The accuracy plots first increase as the increment of clean accuracy. However, the prolonged simulation time improves the approximation of floating-point ANN activation and hampers the robust characteristics caused by the discretization of spike timing. In short, the choice of simulation time is critical for the robustness of rate-based converted SNNs. For example, for the MNIST dataset, the converted SNN running for 9ms is 6.8\% more robust than the ANN when attacked with PGD-$l_2$ of an intensity of 0.5. To step further and analyze the underlying cause behind robustness, we calculated the distance ($l_2$ and $l_\infty$) between the firing rate of the attacked SNN and the attacked ANN activation of the hidden neurons, and the distance between the firing rate of the attacked SNN and the clean SNN of the hidden neurons (Fig.~\ref{fig:part1}(E)). We found that in the process of SNN getting more robust, the fact is not affected that the firing rates of SNN approach ANN activations. The main difference lies in the relationship between clean and attacked SNNs. The distances between firing rates first increase by the low-resolution quantization of the injected noise. This increasing trend can help explain why the converted model is more robust at low simulation times.

Poisson and Current codings encode information in the time window uniformly, making the output of SNNs and ANNs closer after long simulations. Thus, SNNs will have a hard time exhibiting distinction. Here we designed a special deterministic rate coding with synchronization (abbreviated as Rate-Syn): the average firing rate corresponds to the 0-1 floating-point input pixel intensity, and all spikes will be tightly arranged within a time window, and these windows will end synchronously (refer to Fig.~\ref{fig:part1}(B)). We refer to Supplementary Fig.~S1 for a visualization of the three coding schemes. Fig.~\ref{fig:part1}(D) illustrates that the attack accuracy with regard to simulation time is significantly higher than that of Poisson and Current codings. The Rate-Syn coding achieves 96.41\% accuracy under PGD-$l_2$($\varepsilon=0.5$) attack. Note that in the last few time steps, the accuracy suffers from a little decrease. The distances plots in Fig.~\ref{fig:part1}(E) can help explain what is happening. It is shown that using Rate-Syn coding significantly decreases the $l_2$ and $l_\infty$ distances between clean firing rates and attacked firing rates, which are previously assumed to be responsible for the robustness of rate-based SNN. Intuitively, compared to Poisson coding, Rate-Syn coding will postpone the spike timing. Thus, we collected and calculated the instantaneous firing rate of the hidden neurons populations under and without attack plotted in Fig.~\ref{fig:part1}(F), and found that the Rate-Syn coding has a lowered population rate initially and, indeed, needs more time to make up the overall firing rate, and the firing rate for the Rate-Syn coding continuously increase. When the encoding time reaches an end, the robustness of Rate-Syn under attack will decrease.

\subsection{Dedicated training algorithms help spiking neural networks to further improve robustness.}

\begin{figure}[t]
\begin{center}
\centerline{\includegraphics[width=1.0\columnwidth]{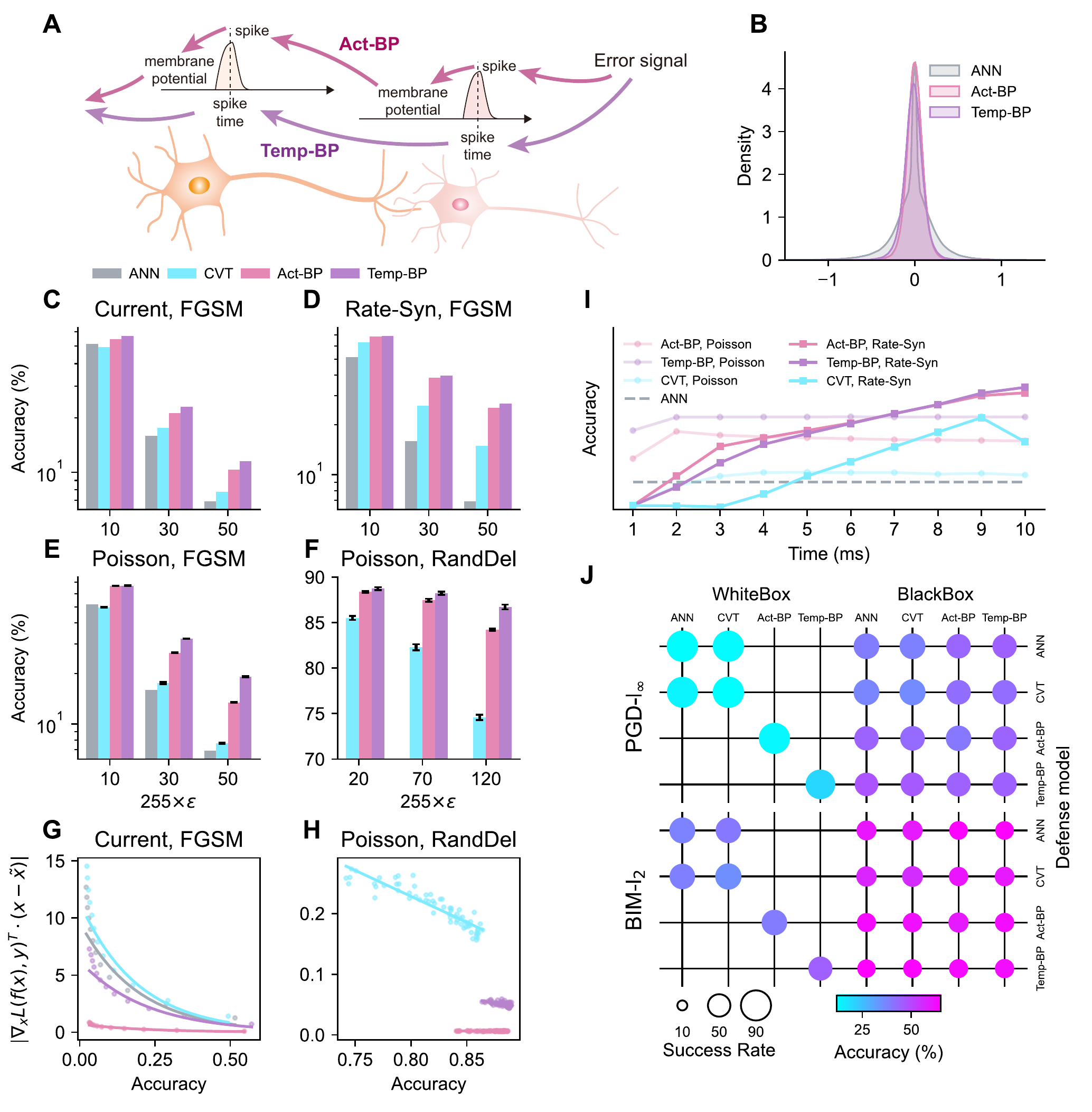}}
\caption{Impact of training algorithms on the robustness of SNN. (A) Diagram of the Act-BP and Temp-BP training algorithms. (B) Weight distribution of Act-BP, Temp-BP of SNN and that of vanilla BP for ANN after training. 
Model accuracy under attack (see Supplementary Fig.~S2 for details): (C) Using FGSM attack for Current coding; (D) Using FGSM attack for Rate-Syn coding; (E) Using FGSM attacks for Poisson coding; (F) Attacking input spikes with random deleting for Poisson coding. (G and H) First-order Taylor polynomial with regard to the model accuracy. The scatters represent results from different attack intensity. (I) Accuracy under FGSM attack with regard to simulation time. (J) Results of ANN, converted SNN (CVT), Act-BP SNN, and Temp-BP SNN performing white-box and black-box attacks (see Supplementary Tab.~S1 for details). }
\label{fig:part2}
\end{center}
\end{figure}

From the above analysis, ANN can improve robustness through conversion to SNN. The conversion-based method normally depends on non-leaky IF neuron. This 's not all that SNNs bring to robustness: Neuron models with leaky factors have been proven to process more robustness due to the leaky effect~\citep{sharmin2019comprehensive}. How to train these spiking neural networks have now become a hot research topic~\citep{neftci2019surrogate, fang2021incorporating, yin2023accurate}. The training methods accurately manipulate on the sequence of spikes, which can be largely divided into two categories: activation-based back-propagation (Act-BP)~\citep{wu2018spatio} and temporal-based back-propagation (Temp-BP)~\citep{zhang2020temporal, kheradpisheh2020temporal}. The main discrepancy of the two categories lies in the procedure of the gradient propagation (see Fig.~\ref{fig:part2}(A)). The Act-BP training smooths the non-differentiable spikes and pass the gradients through the membrane potential $V$, while the Temp-BP training propagate the gradients to the spike time $t$ instead. We followed Wu et al.~\citep{wu2018spatio} and Zhang et al.\citep{zhang2020temporal} to implement Act-BP and Temp-BP, respectively. The weight distributions brought about by the two training algorithms are indeed different~(Fig.~\ref{fig:part2}(B)). We trained SNN with Current, Poisson, and Rate-Syn coding on the Fashion MNIST dataset with Act-BP and Temp-BP and compared the robustness with that of converted (CVT) SNN and ANN. We first tested the effect of black-box FGSM attack. The performance using the three input encoding methods can be seen from Fig.~\ref{fig:part2}(C)(D)(E). In these three experiments, the order of robustness we roughly observed from high to low is: Temp-BP SNN, Act-BP SNN, CVT SNN, and, ANN. The results show that the input encoding also affects the robustness, despite changing the training method. 
To further understand the impact of the input encoding, we plotted the change in accuracy under FGSM attack within 10ms in Fig.~\ref{fig:part2}(I). Note that the synchronization time of Rate-Syn coding is also 10 ms. We found that for Poisson coding, the robustness of shorter time is almost unaffected. And for Rate-Syn, as the encoding progresses, its performance continues to improve and eventually exceeds the performance of Poisson encoding. 

Perception of spike sequence in training improves the stability after perturbation. Surprisingly, we found that the robustness of SNNs trained by Act-BP and Temp-BP is less susceptible to spike time changing compared to CVT SNN. We randomly dropped spikes in the Poisson pattern at some rate~(Fig.~\ref{fig:part2}(F)). Temp-BP SNN and Act-BP SNN also outperform CVT SNN. 

To understand the influence of training methods beyond specific attack methods, we took a deeper look at the characteristics of gradients. We inspected the norm of the gradient $\nabla_{\boldsymbol{x}}\mathcal{L} (f(\boldsymbol{x}),y)$ multiplied by $(\boldsymbol{x} - \tilde{\boldsymbol{x}})$ where $\tilde{\boldsymbol{x}}$ is the attacked sequence, and $y$ is the target. $\lvert \nabla_{\boldsymbol{x}}\mathcal{L} (f(\boldsymbol{x}),y)^T(\boldsymbol{x} - \tilde{\boldsymbol{x}}) \rvert $
contains information of the first-order Taylor polynomial though the gradient of SNN is not accurate. A smaller value indicates the better robustness. We observed from Fig.~\ref{fig:part2}(G)(H) that the norms for CVT SNN and ANN are larger than those of Temp-BP SNN and Act-BP SNN, which may help partially explain the robustness.

Researchers proposed that SNN attacks can also be performed when using surrogate gradients~\citep{kundu2021hire}. We used two groups of models with Current coding to act as defense and attack models, and conducted PGD-$l_\infty(\varepsilon=0.1)$ and BIM-$l_2(\varepsilon=0.5)$ attacks. The model accuracy and attack success rate under attack are shown in Fig.~\ref{fig:part2}(J). 

For the white-box attack, we found that the white-box attack of the naturally differentiable ANN is more powerful than the generated attack by Act-BP and Temp-BP. At the same time, we found that Act-BP and Temp-BP also have weaker black-box attack capabilities. When the ANN model is converted to SNN, its attack capability does not change much. Most importantly, we found that black-box attacks from Temp-BP and Act-BP SNNs towards ANN are weaker than black-box attacks from ANN towards Temp-BP and Act-BP. The unique gradient approximation method of SNN makes SNN have stronger resistance to attack and weaker attack ability.

\subsection{Diverse combinations of neural encoding and decoding criteria enhance the robustness of SNNs.}

\begin{figure}[t]
\begin{center}
\centerline{\includegraphics[width=1.0\columnwidth]{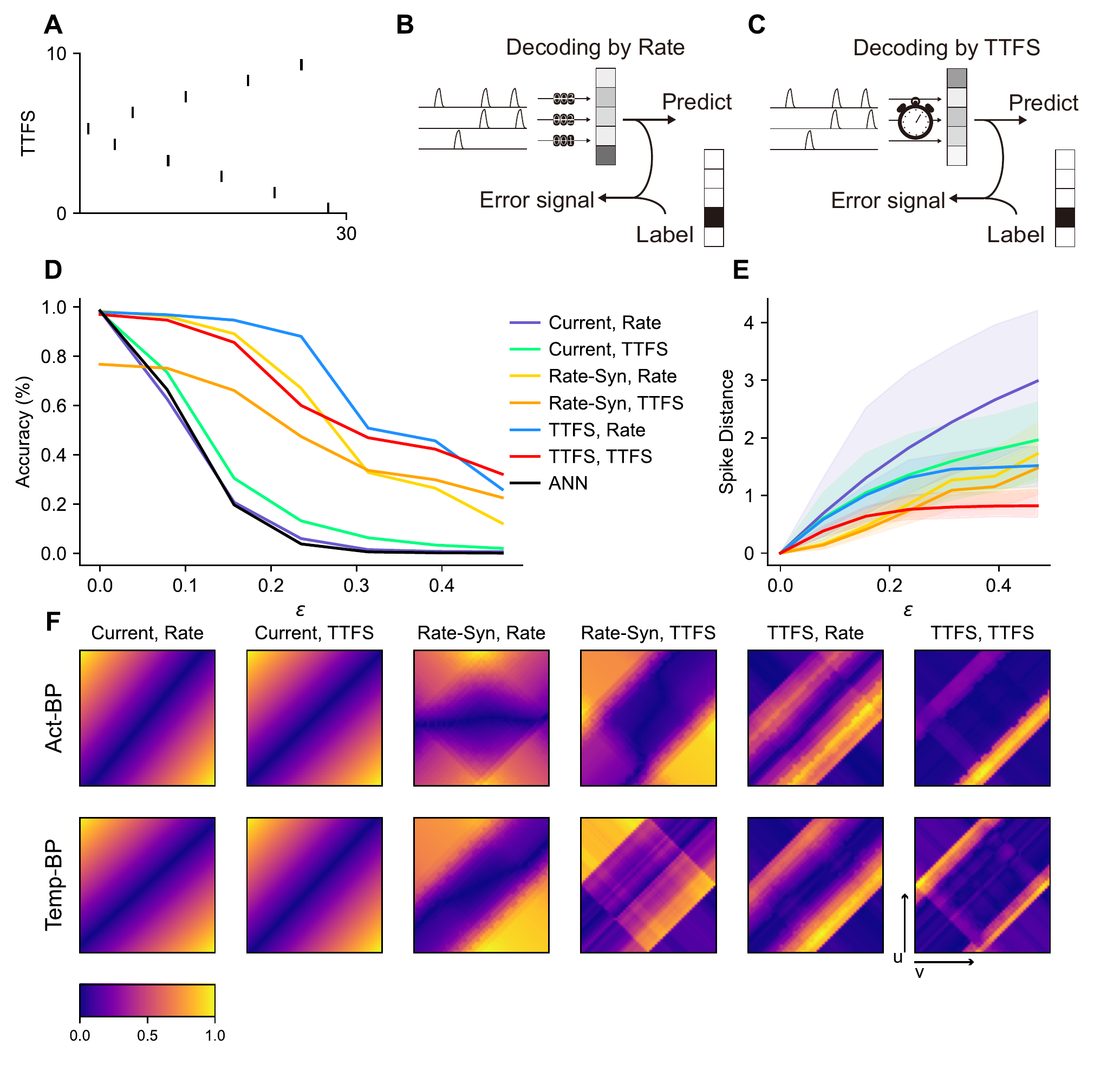}}
\caption{Impact of combinations of encoding and decoding for SNN. (A) Illustration of the TTFS encoding. (B) Illustration of the Rate decoding. (C) Illustration of the TTFS decoding. (D) Accuracy under FGSM attack for the combinations of encoding and decoding. (E) Change of spike distance with regard to the attack intensity. (F) Heatmap of the first-order Taylor polynomial $\vert \nabla_{\boldsymbol{x}}\mathcal{L} (f(\boldsymbol{x}),y)^T (\varepsilon _1\vec{\boldsymbol{u}}+\varepsilon _2\vec{\boldsymbol{v}}) \vert$, where $\vec{\boldsymbol{u}},\vec{\boldsymbol{v}}$ are two adversarial directions, and $\varepsilon _1$ and $\varepsilon _2$ are scalars in $\left [0,1\right]$.}
\label{fig:part3}
\end{center}
\end{figure}

We already know that an increase in the robustness of SNNs can be achieved by incorporating the SNN-specific approximation gradients into training. In addition, the gradient can also be varied by adjusting the strategy of decoding. Currently, the most commonly used decoding strategy is through firing rate, that is, determining the model prediction from the magnitude of the firing rate of output neurons~\citep{lee2020enabling, sengupta2019going}. Besides, the precise spike time delivered by temporal coding contains rich information and has also been used to provide an error signal for decoding~\citep{kim2020unifying}. In order to test the influence of temporal coding on SNN, TTFS, a typical temporal coding strategy, is used for encoding and decoding in SNNs. As illustrated in Fig.~\ref{fig:part3}(C), decoding by TTFS means the prediction and error signal focus on the latency of the first spike (abberivated as TTFS decoding), rather than firing rate (abberivated as Rate decoding). We used Current coding, Rate-Syn coding, and TTFS coding as input encoding, and Rate decoding and TTFS decoding as decoding methods. The combination of encoding and decoding methods are accompanied with Act-BP to train SNN on the MNIST dataset: Current+Rate, Rate-Syn+Rate, TTFS+Rate, Current+TTFS, Rate-Syn+TTFS, TTFS+TTFS.

We found that the robustness varies for the combinations in our experiments. We put the six well-trained SNNs and ANN under the attack of black-box FGSM (Fig.~\ref{fig:part3}(D)). For Current and Rate-Syn coding, TTFS decoding helps improve robustness. At the same time, the robustness of the model using TTFS encoding is high, no matter which decoding method is used. In short, the addition of temporal encoding and decoding makes the model more resistant to black-box FGSM attacks compared to ANN. The performances of these methods under PGD attack are similar, as shown in Supplementary Fig.~S3.

We attempted to explain the reason for the robustness from two perspectives. First, we recorded the spike trains produced by neurons in the hidden layer receiving clean and attacked inputs as the attack strength increasing and calculated their Victor-Purpura distance (Fig.~\ref{fig:part3}(E)). We found that models with higher perturbation immunity generally have smaller spike distances. In addition, we randomly selected two mutually orthogonal FGSM perturbation directions ($\vec{\boldsymbol{u}},\vec{\boldsymbol{v}}$) and obtained the first-order Taylor polynomial 
$\Vert \nabla_{\boldsymbol{x}}\mathcal{L} (f(\boldsymbol{x}),y)^T (\varepsilon _1\vec{\boldsymbol{u}}+\varepsilon _2\vec{\boldsymbol{v}}) \Vert$, where $\varepsilon _1$ and $\varepsilon _2$ are scalars in $\left [0,1\right]$. We presented these results in the heatmap in Fig.~\ref{fig:part3}(F). It is found that for the less robust combination Current+Rate, it always has a larger value of the first-order Taylor polynomial, which means that when a perturbation is applied, the combination is easily affected. In contrast, the other combinations have larger `valleys' in the heatmap. This means that there are quite a few perturbations that do not bring about large changes in the first-order Taylor polynomial. For the two combinations of TTFS+Rate and TTFS+TTFS, which have strong robustness in our experiments, the heat map is covered almost by `troughs'. We also plotted the results of SNNs trained by Temp-BP and got similar inductions. 

\subsection{Application to real-world scenarios.}

\begin{figure}[t]
\begin{center}
\centerline{
    \includegraphics[width=1.0\columnwidth]{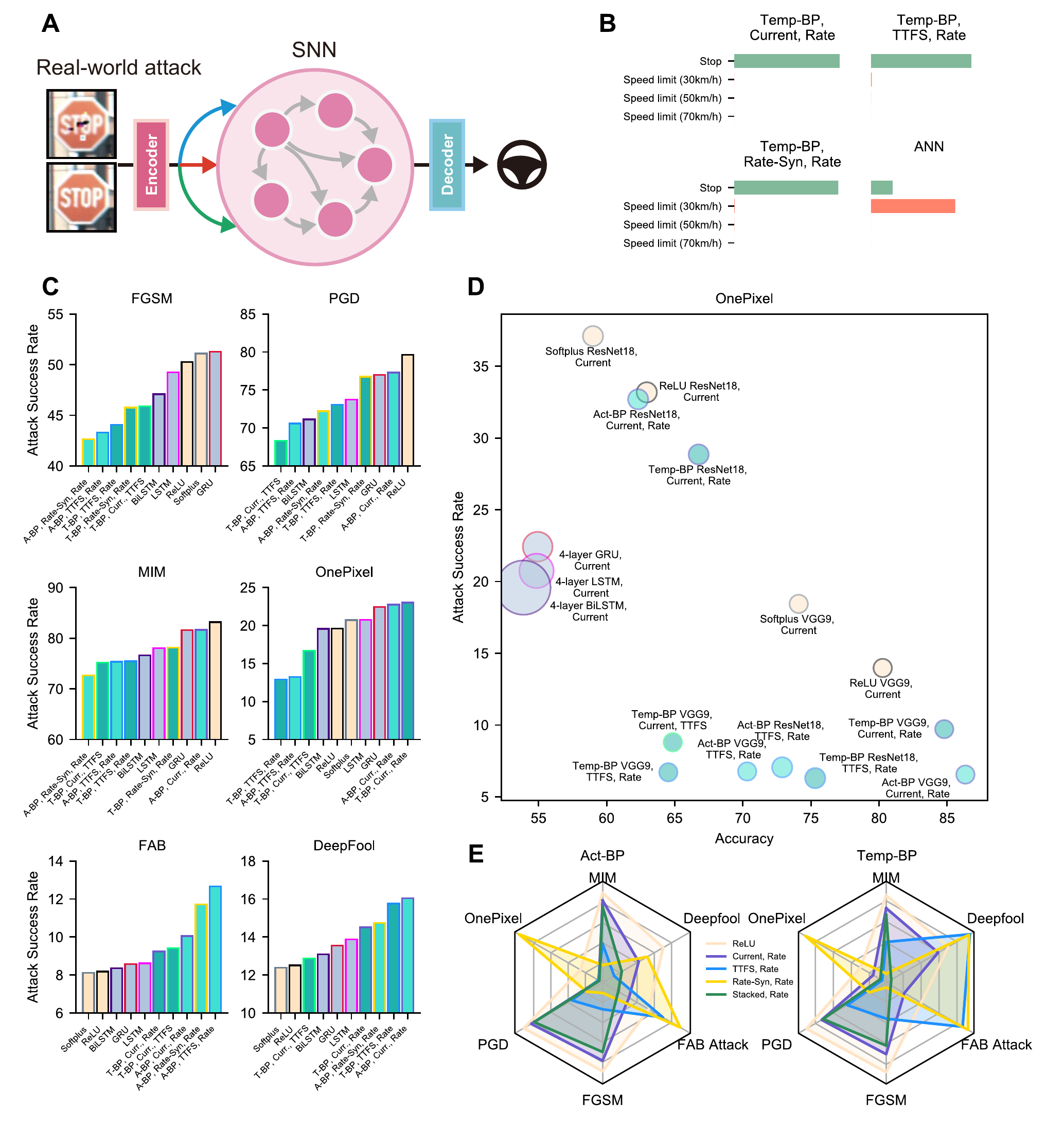}
    }
\caption{Application of SNN to real-world scenarios. (A) Illustration of SNN making decisions in the scenario of real-world attacks. The attack can be conducted by stickers, graffiti and so forth. (B) Predictions made by SNN and ANN of the attacked stop sign in (A). (C) Attack success rate under black-box attacks. The attack methods include FGSM, PGD, MIM, OnePixel, FAB, and DeepFool attacks (see Supplementary Tab.~S2 for details). (D) Attack success rate and accuracy under OnePixel attack. The size of the scatters corresponds to the amount of model parameters. (E) Attack success rate of the stacked SNN under multiple attack methods (see Supplementary Tab.~S2 for details).}
\label{fig:part4}
\end{center}
\end{figure}

We discussed the impact of input coding, output decoding, and training methods of SNNs on robustness before. Even for black-box attacks, it is difficult to achieve pixel-level perturbation in real-world scenarios. The more likely way is to apply real-world attacks, such as manually designed stickers, graffiti~\citep{eykholt2018robust}, laser~\citep{duan2021adversarial}, reflection~\citep{liu2020reflection} and other means. We wondered whether the SNNs can resist more attack methods and performs robustly in daily scenarios. Therefore, we conducted experiments on the German Traffic Sign Recognition Benchmark (GTSRB) dataset. The GTSRB dataset contains images of 43 types of traffic signs that are common in daily life. 

We trained SNN on classification tasks for simulating a safety-critical application on vehicles. We first followed the method proposed in \cite{eykholt2018robust} to conduct a black-box attack on the stop signs in the test data set. As a result of the attack, patterned stickers appeared on these samples. These adversarial examples are fed into the SNN through the encoder, and the decoding result of the output sequence affects the decision of the system~(Fig.~\ref{fig:part4}(A)). 

We compared ANNs and SNNs with various combinations of settings at the same network size. Experiments showed that compared to the 27.8\% accuracy obtained by the ANN with ReLU activation, the Temp-BP SNN under the Current+Rate setting accurately identified 55.6\% of the attacked samples. For the adversarial example in Fig.~\ref{fig:part4}(A), when ANN recognizes the stop sign, it recognize the sign as `Speed limit (30km/h)', and this sample do not confuse the SNN under certain settings~(Fig.~\ref{fig:part4}(B)). 

At the same time, we also tested the robustness of these SNNs on more perturbation methods. We trained a ReLU-activated ResNet18 and used this model to construct black-box attack samples. The attack methods include FGSM~\citep{goodfellow2014explaining}, PGD~\citep{aleksander2018towards}, MIM~\citep{dong2018boosting}, OnePixel~\citep{su2019one}, FAB~\citep{croce2020minimally}, and DeepFool~\citep{moosavi2016deepfool}. We also trained Softplus-activated ANN as well as recurrent neural network (RNNs) with gated recurrent units (GRU) and long short-term memory (LSTM) for comparison. These models also have the same neuron scale. We compared the success rate of these attacks. For the defense model, a lower attack success rate means better robustness. We sorted the results of the top ten models with the lowest attack success rate and displayed them in Fig.~\ref{fig:part4}(C). 

For the OnePixel attack, we tested the performance of deeper SNNs and plotted a scatter plot (Fig.~\ref{fig:part4}(D)). Note that the size of the scatters indicates the amount of model parameters. Specifically, we experimented with VGG9 and ResNet18 network architectures. The increase of model depth does not improve the attack success rate. For some SNNs trained by Temp-BP and Act-BP, due to the improvement of the generalization ability brought by the deeper architectures, the accuracy of the adversarial samples is also improved, and the attack success rate is greatly reduced, such as for VGG9 SNN trained by Act-BP.

We considered that the robust properties of SNNs are not exactly the same for different coding and decoding combinations, and that the brain does not use a single encoding for signal transmission. Therefore, we used the stacking technique to integrate the VGG9 SNNs that uses Rate decoding~\citep{wolpert1992stacked}. The stacking technique obtains features from three models of different coding, and concatenates the features. The new features are used to retrain a linear SNN layer and give Rate-decoded prediction. We performed ensembles on Act-BP and Temp-BP SNNs separately and tested the performance of the stacked models. We displayed the attack success rates of these models against six adversarial methods in radar charts (Fig.~\ref{fig:part4}(E)). Any single coding scheme only provides better resistance to some attacks compared to the ReLU-activated ANN. However, the stacked model provides better robustness to any kind of chosen attack method, which suggests the benefit of using a fused coding strategy in SNN.

% ~\\
% \noindent \textbf{XXX XXX XXX XXX.} XXX XXX XXX

\section{Discussion}

We have systematically demonstrated in this paper that precise spike timing is conducive to improving the robustness of neural networks, providing opportunities for understanding the robustness of the brain. We gave a feasible path of Securing Artificial Intelligence (SAI) from the computing point of view, that is, to realize brain-inspired deep learning by using SNN and fully adopting precise time encoding and decoding. In particular, we found that robustness can be improved simply by adjusting the synchronicity of the neural encoding, and this advantage is evident when using SNN-specific training methods. Moreover, combined with various decoding schemes, we can achieve robust training with a perception of the loss landscape. Furthermore, when these schemes form hybrid systems, we demonstrated the robustness of the hybrid model to a wide variety of attack methods.

SNNs are an evolved version of ANNs~\citep{maas1997networks}, which naturally extend timing capabilities to neural networks~\citep{rao2022long}. However, this timing capability has not been well utilized so far. Neural network robustness is an excellent showcase to illustrate the importance of timing capability. In particular, we designed a concise time encoding method that can improve the robustness of the SNN models converted from ANNs only by regulating the start and end firing times of input neurons. Previous work also found that coding is important for the robustness of SNNs~\citep{kundu2021hire, kim2022rate}, but stayed in the discussion of spike count rate coding. In fact, the robustness gain that the spike count rate coding can bring through discrete-time quantization is very limited, as demonstrated in the current study. Therefore, we designed precise input encoding strategies to achieve neuronal synchronization-like functions. This scheme works when using weights converted from ANN. This inspires us to re-understand ANN: ANN does not lack defense ability but adopts an ineffective input scheme. Given neuronal synchronization has a broad biological basis~\citep{womelsdorf2007modulation}, and temporal multiplexing of neural coding is efficient for information representation~\citep{Panzeri2010}, we hope that our model can help explain more complex biological functions.

Since we adopted a more complex SNN encoding scheme, the specific training scheme of the SNN needs to be considered to assist in performance improvement. The change in the training method enables SNN to better utilize the advantages of timing. One of the recent concentrations in SNN is how to design a better training algorithm for SNN. The Act-BP and Temp-BP mentioned in the paper are among these algorithms. Act-BP and Temp-BP represent two types of algorithms. The backpropagation process of Act-BP models the complete dynamic process of spike generation~\citep{lee2020enabling, wu2018spatio}, while Temp-BP pays more attention to how to use the sparsity of spike timing during training~\citep{zhang2020temporal, zhu2022training}. In the experiments, we found that both training methods improved the robustness of SNNs. This suggests that accurate spike timing encoding requires the assistance of SNN training methods, and work on how SNNs are trained is critical to exploiting temporal information. Of course, the training method of SNNs also drives the research on SNN-specific adversarial methods~\citep{liang2021exploring}. In our work, we considered the ability of the network trained by these different methods to carry out attacks. The model with more robustness fails to provide better attack capabilities. Our results could inspire a series of studies on SNN attack and defense to overcome the challenges that ANN is currently facing~\citep{davies2019benchmarks}.

After adopting SNN-specific learning methods, it becomes possible for SNN to utilize different neural encoding and decoding methods. Neural decoding is concerned with how neuronal responses are translated into meaningful labels~\citep{nakai2022representations, suarez2021learning}. The most common decoding scheme in image recognition is to use the average firing rate over time as a vector for model predictions~\citep{cao2015spiking}. Similar to encoding, this actually ignores the positive effect of spike timing. To this end, we integrated the timing-sensitive TTFS decoding scheme. Accompanied by the adjustment of the decoding scheme is the change of the loss function, which alters the SNN's perception of the loss landscape. Comparing rate decoding and TTFS decoding, we found that they correspond to discrepances in the first-order Taylor polynomial landscape near the input. This may be due to the fact that the loss function constructed based on TTFS decoding is more sparse. Therefore, it can inspire research to develop more robust decoding strategies. At the same time, TTFS also participated in the experiment as an encoding scheme, and we found that this sparse code can also bring robustness.

SNN is beneficial for being implemented on various edge devices with a low energy budget~\citep{wozniak2020deep, frenkel2021sparsity}, but real-world scenarios are more complex and need to deal with inputs of different scales and perturbations of different degrees~\citep{komkov2021advhat}. Compared with the FGSM attack, there are more complex and sparse perturbations from only changing a few pixels~\citep{su2019one}. We used the dataset to simulate the results of real pictures under different perturbation conditions. SNNs are capable of providing highly robust solutions to various perturbations. Nevertheless, we did not find a certain combination of SNN configurations to provide robustness to perturbations in all experiments. Complex coding schemes are employed in the visual pathway~\citep{webster2011adaptation}. In our experiments, the SNN trained under the fused encoding could compensate for the robustness of various attacks, proving the necessity and effectiveness of the fused encoding. These results could help explain the role various neural coding systems play in biological systems.

At present, our experiment does not include any defense measures against perturbations. We are only exploring the robustness brought by SNN itself using spike timing. The lack of active defense measures is a limitation of our research. Our discussion of SNNs currently focuses on point neurons without considering the impact of complex neuronal structures and processes (such as synaptic conduction) on precise spike timing. This limitation is actually brought about by the SNN training technique. The current mainstream training scheme is mainly applicable to SNN constructed by point neurons. In addition, these training algorithms also limit the increase in simulation time step size, so the robustness of SNNs under longer time steps remains to be explored. Besides, we mainly used the black-box gradient attack obtained by the ANN, which is naturally differentiable. On the one hand, it is to consider the fairness of the experiment. On the other hand, SNN-specific attack methods are still immature. The robustness of SNNs under these potential attacks is still an open question for future studies. Our work here provides a first systematical investigation of this question, paving the way for looking into the role of each component, encoding, decoding, and learning, of SNNs in next-generation brain-inspired computing models. 

\section{Methods}

\subsection{Neuron and synaptic models}
Spiking neural networks (SNNs) are third-generation neural networks that follow the principle of neuronal dynamics that originated in biological neurons. We compared SNNs mainly with artificial neural networks (ANNs), the most popular type of neural network in machine learning. ANNs normally consist of multi-layer neurons that have a non-linear activation function. In this work, we mainly refer to two activation functions: ReLU ~\citep{fukushima1975cognitron} and Softplus~\citep{zheng2015improving}.
\begin{equation}
\label{eq:relu}
    \text{ReLU}(\bm{x}) = \max(\bm{x},0),
\end{equation}
\begin{equation}
\label{eq:softplus}
    \text{Softplus}(\bm{x}) = \log(1+\exp(\bm{x})),
\end{equation}
where $\bm{x}$ is the vector of the input. Note that the ReLU activation is used mostly for ANN, and in the fourth part of the results, we applied the Softplus activation. 

SNN uses spiking models to process information. The leaky and non-leaky integrate-and-fire model (IF, LIF) are common types of spiking models that are deployed in this work~\citep{gerstner2014neuronal}. The membrane potential of neuron $i$ in the layer $l$ varies over time as Eqs.~\ref{eq:if},\ref{eq:lif}.
\begin{equation}
\label{eq:if}
    \text{IF:}~\frac{dU_i^{(l)}(t)}{dt} = R I_i^{(l)}(t),
\end{equation}
\begin{equation}
\label{eq:lif}
    \text{LIF:}~\tau_m \frac{dU_i^{(l)}(t)}{dt} = -(U_i^{(l)}(t) - U_0) + R I_i^{(l)}(t),
\end{equation}
where $\tau_m$ is the membrane time constant, $R$ is the resistance, $I_i^{(l)}(t)$ is the input current, $U_i^{(l)}(t)$ is the membrane potential, and $U_0$ is the resting potential. Eq.~\ref{eq:if},\ref{eq:lif} can be all reduced to $\frac{dU_i^{(l)}(t)}{dt} = f(U_i^{(l)}(t), I_i^{(l)}(t))$. Considering the discrete implementation in the computer, the fixed-step first-order Euler method is used to discretize the dynamic functions (Eq.~\ref{eq:discrete}). $\Delta U_i^{(l)}[t]$ and $U_i^{(l)}[t]$ take the place of $\frac{dU_i^{(l)}(t)}{dt}$ and $U_i^{(l)}(t)$.
\begin{equation}
	\label{eq:discrete}
	\begin{split}
	\Delta U_i^{(l)}[t] &= f(U_i^{(l)}[t], I_i^{(l)}[t]),\\
	U_i^{(l)}[t+1] &= \Delta U_i^{(l)}[t] + U_i^{(l)}[t].\\
	\end{split}
\end{equation}

When the membrane potential $U_i^{(l)}[t]$ reaches the threshold $U_{th}$, the spiking neuron generates a discrete spike from Heaviside step function $H(\cdot)$ and then $U_i^{(l)}[t]$ is driven back to $U_0$. The spikes emitted can be formed into a spike train as in Eq.~\ref{eq:spike_train}.
\begin{equation}
	\label{eq:spike_train}
	S_i^{(l)}[t] = \sum_k \delta(t-t_{i}^{(l),k}).
\end{equation}
The postsynaptic potential $a_i^{(l)}[t]$ simulates a temporary change by applying the spike response kernel function. Here we adopted a first-order synaptic model for $\varepsilon(\cdot)$.
\begin{equation}
	\label{eq:psc}
	a_i^{(l)} = (\varepsilon * S_i^{(l)})(t),
\end{equation}
where $\tau_s$ is a synaptic time constant in $\varepsilon(\cdot)$. Note that the trained SNNs are LIF SNN. We listed the default parameters of the single neuron experiment in Table~\ref{tab:default_neuron} if not specially noted.

\begin{table}[t]
	\centering
	%\tiny
	\caption{Default parameter for neuron models.}
	\label{tab:default_neuron}
	\begin{tabular}{ccc}
		Parameter  & Value & Description                  \\ \hline
		$R$        & 1.0   & Resistance                   \\
		$\tau_m$   & 2.0   & Membrane Time Constant       \\
            $\tau_s$ & 2.0   & Synaptic Time Constant            \\
		$U_0$     & 0.0   & Resting Potential            \\
		$U_{th}$   & 1.0   & Membrane Threshold           \\
	\end{tabular}
\end{table}

\subsection{Encoding and decoding schemes}

To feed images into SNN, input encoding is the first consideration. We implemented four input coding schemes in this work, namely Current coding, Poisson coding, Rate-Syn coding, and TTFS coding. Current coding here represents a method where the floating-point pixel matrix of images is directly fed into SNN, lasting for some timesteps. In the literature of SNN, it is also referred to as Direct coding~\citep{kim2022rate}. Assume that $X$ is the image pixel intensity, it can be formulated in Eq.~\ref{eq:current_code}.
\begin{equation} 
\label{eq:current_code}
    I_{Current}[t] = X, \,\,\,   t=1,2,\cdots,T,
\end{equation}
where $T$ is the duration of input. 

Poisson coding, as the most prominent branch of rate coding, is a typical candidate of rate coding and processes a neuronal background~\citep{amarasingham2006spike}. It is widely used in the field of SNN~\citep{maass1999computing, gabbiani1998principles}. It is also first applied to connect SNN with ANN by conversion~\citep{diehl2015fast}, which can be formulated in Eq.~\ref{eq:poisson_code}.
\begin{equation} 
\label{eq:poisson_code}
    I_{Poisson}[t] = Rand(0,1) < X,  \,\,\,   t=1,2,\cdots,T,
\end{equation}
where $Rand(0,1)$ is a random generator that generates uniformly distributed values for each item of the pixel matrix in the range of 0 and 1. 
Both Poisson and Current coding produce a steady firing rate that is categorized as Count Rate coding~\citep{auge2021survey}. These methods encode pixel intensity in the mean firing rate. However, it is observed that both coding schemes have a limited increase in robustness~(Fig.~\ref{fig:part1}). 

To break the steady instantaneous firing rate and maintain the mean firing rate, we developed the Rate-Syn coding scheme. Rate-Syn coding is short for rate coding with synchronization, which we design here to illustrate that using typical rate coding is not enough to carry much robustness. This coding scheme can be expressed in Eq.~\ref{eq:ratesyn_code}.
\begin{equation} 
\label{eq:ratesyn_code}
    I_{Rate-Syn}[t]=\begin{cases}
	1,\,\,\,t\ge (1-X)T,t=1,2,\cdots,T\\
	0,\,\,\,t<(1-X)T,t=1,2,\cdots,T\\
\end{cases}
\end{equation}
The spike trains generated by the Rate-Syn coding have a synchronized ending time. The time when the first spike starts depends on the pixel value. The mean firing rate during T steps approximates one of Poisson and Current coding. 

Apart from considering count rate coding, a basic coding scheme, TTFS coding (Time-to-first-spike coding) is also taken into consideration. In TTFS coding, the first spike time encodes the information, which can be formulated in Eq.~\ref{eq:ttfs_coding}.
\begin{equation}
\label{eq:ttfs_coding}
    I_{TTFS}[t] = 1, \,\,\, t = (1-X)T.
\end{equation}
The difference between Rate-Syn and TTFS codings lies in the number of spikes. TTFS coding offers one spike in a single spike train, while Rate-Syn continues to fire since $(1-X)T$. 

Count rate coding and temporal coding represent the two forms of transformation from pixel value to spike train. The prediction of SNN depends on the manipulation (or decoding) from the spike train to digits. Similar to the design of coding methods, we used two decoding methods in this work, that is Rate decoding and TTFS decoding. The rate decoding translates a spike train to the mean firing rate, while TTFS decoding allows more than one neuron only focuses on the first spike time and discards the rate information. The decoding methods also influence the design of losses. When training the network, the decoded vectors are targeted to the true labels.

\subsection{Architectures and Training approaches}

For Figs.~\ref{fig:part1}, \ref{fig:part2}, \ref{fig:part3}, to avoid the robustness influence brought by complex architectures, we experiment on the fully connected network with no bias both for ANNs and SNNs. Our experiments are conducted on either MNIST or Fashion MNIST dataset. The network has three feed-forward layers. The input neuron number is 784 (28$\times$28), which matches the number of pixels from an image of MNIST or Fashion MNIST dataset. The model outputs probabilities of 10 classes to predict. We decoded the SNN output with the mean firing rate by default. 

For ANN, we optimized the model using backpropagation with adaptive momentum estimation optimizer with weight decay (AdamW)~\citep{loshchilov2017decoupled,kingma2015adam}. The decay factor is set to $0.01$, and the learning rate is $0.001$. The loss used to train ANN is cross-entropy loss, and the model is trained for 100 epochs without further indications.

Three training approaches of SNN are considered in this work. They are conversion-based approaches (CVT), activation-based back-propagation (Act-BP), and temporal-based back-propagation (Temp-BP). We denote the synaptic weights as $W_{ij}^{(l)}$ for layer $l$ here.

First, high-accuracy SNN can be obtained from converting and rescaling the weights of ANN. The fundamental principle of conversion is to match ReLU nonlinearity with the firing rate of IF neurons. At the moment of a spike, the membrane potential $U_i^{(l)}[t]$ is reduced by an amount equal to the firing threshold $U_{th}$, instead of going back to the resting potential. The exact conversion restricts the use of postsynaptic potential, and only spikes are transferred. We rescaled the weights following the methods proposed by \cite{rueckauer2017conversion}. Given the maximum activation of layer $l$ and $l-1$ in ANN as $\max^{(l)}$ and $\max^{(l-1)}$, the weights are scaled following Eq.~\ref{eq:CVT}.
\begin{equation}
	\label{eq:CVT}
	W_{ij}^{(l),SNN} = W_{ij}^{(l),ANN} \cdot U_{th} \cdot \frac{\max^{(l-1)}}{\max^{(l)}},
\end{equation}
where $W^{(l),SNN}$ is the converted weight in SNN and $W^{(l),ANN}$ is the raw ANN weight. The conversion-based SNN needs a larger time steps to rival the raw ANN in accuracy.

Conversion-based approaches do not need backpropagate through SNN. Act-BP and Temp-BP methods have been proposed to overcome this problem. These methods can make LIF neurons better trained, so we use Act-BP and Temp-BP to train the model of LIF neurons. The forward pass of layer l in LIF SNN can be discretized in Eq.~\ref{eq:forward_snn}.
\begin{equation}
	\label{eq:forward_snn}
	\begin{split}
		U_i^{(l)}[t+1] &= (1 - \frac{1}{\tau_m})U_i^{(l)}[t] + \sum_j W_{ij}^{(l)} a_j^{(l-1)}[t]\\
		S_i^{(l)}[t] &= H(U_i^{(l)}[t] - U_{th}) \\
		 a_i^{(l)}[t+1] &= (1 - \frac{1}{\tau_s}) a_i^{(l)}[t] + S_i^{(l)}[t], \\
	\end{split}
\end{equation}
where $\tau_s$ is synaptic time constant and $H(\cdot)$ is the Heaviside step function.

For Act-BP methods, the main idea is to smooth the Heaviside step function. To do this, we adopted the sigmoid surrogate function as in \cite{neftci2019surrogate} so that the non-differentiable function can have a gradient that can be used to update paramenters. In the forward pass, the network follows a step function as in Eq.~\ref{eq:psc}, and in the backward pass it follows a sigmoid function expressed in Eq.~\ref{eq:surrogate}.
\begin{equation}
	\label{eq:surrogate}
		\frac{\partial S}{\partial U} = \text{Sigmoid}'(\rho \cdot U).
\end{equation}
where the steepness is controled by $\rho$, which is set to 5 by default.

For Temp-BP methods, the key challenge for Temp-BP methods is how to obtain $\frac{\partial a}{\partial U}$ via spike time. Following \cite{zhang2020temporal}, we seperated the $\frac{\partial a}{\partial U}$ into inter-neuron and intra-neuron backpropagation. The inter-neuron backpropagation happens when the postsynaptic potential is triggered by a presynaptic firing time. The intra-neuron dependency is defined between an arbitrary time and a presynaptic firing time. 

For SNN, the learning rate is set to 0.0005, AdamW optimizer is also used and the weight decay is 0.01. All the training is implemented using PyTorch~\citep{paszke2019pytorch}.

In Fig.~\ref{fig:part4}, we demonstrated that the robustness of SNN is maintained for deeper architectures and can be applied to real-world scenarios. We trained SNN on the German Traffic Sign Recognition Benchmark (GTSRB) dataset, which has images of 43 types of traffic signs. The images are first resized to 32 $\times$ 32 with 3 channels. Then we tested three architectures for ANN with ReLU and Softplus activation and SNN with LIF neurons. The first is 
a five-layered fully connected network (3072-1000-1000-1000-1000-43). The other two achitectures are ResNet18~\citep{he2016deep} and VGG9~\citep{karen2015very}. The neuron of these two achitectures is altered to LIF neuron for SNN. To improve generalization on deep networks, we applied a cosine annealing learning rate schedule. We also made comparison on the recurrent networks, that is LSTM~\citep{hochreiter1997long} (bidirectional and not bidirectional) and GRU~\citep{kyunghyun2014on}. These recurrent networks all have the same number of hidden neurons as five-layered networks. 

\subsection{Attack methods}
\label{sec:methods_perturbation}
Various perturbation methods are referred to in this work to verify the robustness. We mainly focus on the adversarial attacks, which pose great threat to the modern artificial intelligence. One type is gradient-based attack methods which need to specify the magnitude of intensity. The process of carrying out attack is solving a constrained optimization problem in Eq.~\ref{eq:attack_type_1}.
\begin{equation}
\label{eq:attack_type_1}
	\arg\max_{\bm{\delta}} \mathcal{L} (f(\bm{x}+\bm{\delta}),y)~~~~s.t.~~\Vert \bm{\delta} \Vert_p \leq \varepsilon,
\end{equation}
where $\bm{\delta}$ is the perturbation for $\bm{x}$, $\varepsilon$ is the intensity constraining the p-norm of $\bm{\delta}$, $\mathcal{L}$ is the loss function of the network. $\bm{x} \in R^{N \times 1}$ where $N$ is the number of pixels. We implemented four methods of this type: fast gradient sign method (FGSM,~\cite{goodfellow2014explaining}), basic iterative method (BIM, ~\cite{kurakin2018adversarial}), projected gradient descent method (PGD,~\cite{aleksander2018towards}), and momentum iterative method (MIM,~\cite{dong2018boosting}).

FGSM is a one-step gradient-based approach that generates attacks within a $l_\infty$ ball:
\begin{equation}
    \bm{\delta} = \varepsilon~\text{sign}(\nabla_{\bm{x}} \mathcal{L} (f(\bm{x}),y)),
\end{equation}
where $\nabla_{\bm{x}} \mathcal{L}$ is the gradient with regard to $\bm{x}$. Instead of using direct gradient, MIM calibrates the gradient estimation by accumulating a momentum in gradient direction. 

For iterative methods like BIM and PGD, they solve optimization problem in an iterative manner with a step size $\alpha$. For $l_\infty$ iterative attacks, the basic iteration can be expressed as:
\begin{equation}
	\widetilde{\bm{x}}^k = \Pi_{l_\infty, \varepsilon} \{ \widetilde{\bm{x}}^{k-1} + \alpha~\text{sign} (\nabla_{\bm{x}} \mathcal{L} (f(\widetilde{\bm{x}}^{k-1}),y)) \},
\end{equation}
where $k$ denotes the number of the iteration step. The data in each iteration should be projected onto the space of the $l_\infty$ ball around clean data $\bm{x}$ with regard to $\varepsilon$.
Similarily, when it comes to $l_2$ iterative attacks, the iteration can be expressed as:
\begin{equation}
	\widetilde{\bm{x}}^k = \Pi_{l_2, \varepsilon} \{ \widetilde{\bm{x}}^{k-1} + \alpha \frac{\nabla_{\bm{x}} \mathcal{L} (f(\widetilde{\bm{x}}^{k-1}),y)}{ \Vert \nabla_{\bm{x}} \mathcal{L} (f(\widetilde{\bm{x}}^{k-1}),y) \Vert^2} \},
\end{equation}
where the projection is on the $l_2$ ball. The initial condition of $\widetilde{\bm{x}}^0$ reflects a big difference for PGD and BIM. BIM uses the raw input as the initial condition: $\widetilde{\bm{x}}^0 = \bm{x}$, while PGD adds Gaussian noises in $\bm{x}$ as $\widetilde{\bm{x}}^0$. In this work, PGD and BIM is projected to both $l_\infty$ and $l_2$ balls to test the robustness against attacks.

With the help of foolbox~\citep{rauber2017foolbox} and torchattacks~\citep{kim2020torchattacks} packages, we performed attacks on the targeted model. When performing attacks, we mainly adopted the setting of black-box (BB) attack, where attackers have no knowledge about the defensive model, considering that SNN have no natural gradient and there are more application scenarios for black-box attacks. For Fig.~\ref{fig:part1}, \ref{fig:part2}, \ref{fig:part3}, attacks are performed from a ReLU-activated ANN trained with a different random seed. The architecture is the same as SNN. For Fig.~\ref{fig:part4}, attacks are performed using a ReLU-activated ResNet18. 
White-box attacks are also conducted in Fig.~\ref{fig:part2}(J). First, for Act-BP and Temp-BP SNNs, the white box attacks utilize the special backpropagation strategy, although not as precise as ANN backpropagation. Besides, for conversion-based SNNs, those SNNs has ANN counterparts so they can perform white-box attacks using ANN counterparts or backpropagating from Act-BP~\citep{sharmin2019comprehensive}. 

In order to show that different training methods can also resist attack to the temporal structure. We specifically designed a random deleting attack (RandDel). The random deleting drop Poisson-distributed spikes. The ratio of the deleted spikes and the total spikes is $p$. By increasing $p$, a stronger attack can be obtained.

Finally, to test the performance of SNN in real application scenarios, we additionally utilize more complex adversarial strategies. These methods do not have an adjustable attack intensity, but try to reduce the intensity and increase the sparsity of the perturbation as much as possible making the neural network give wrong predictions, which can be formalized as a Lagrangian-relaxed problem:
\begin{equation}
    \arg\min_{\bm{\delta }} \lambda \left\| \bm{\delta } \right\| _p-\mathcal{L} (f(\bm{x}+\bm{\delta }),y),
\end{equation}
where $\lambda$ controls the regularization of the perturbation. Such perturbation methods are also known as optimization-based methods. Since the objective directly optimizes the norm of the perturbation, the perturbation is more concealed and easy to operate. We implemented four of this kind: Robust Physical Perturbations (RP2,~\cite{eykholt2018robust}), DeepFool attack~\citep{moosavi2016deepfool}, OnePixel attack~\citep{su2019one}, and FAB attack (Fast Adaptive Boundary,~\cite{croce2020minimally}). For RP2, we used an open-source black-box model to implement the attack. For DeepFool, OnePixel and FAB attack, a ReLU-activated ResNet18 is specially trained to perform attacks.

\subsection{Robustness Analysis}

In this work, we utilized two metrics to evaluate the robustness. One is the drop in clean accuracy. Neural network may give wrong prediction when under attacks, thus bringing about a drop in accuracy. Using a gradient attack algorithm that can adjust the attack intensity and showing the decrease in accuracy as the intensity increases can well characterize the harmful impact of adversarial examples on the network. The more robust the model, the less accuracy drops.

Another metric is attack success rate (ASR), which measures the probability of adversarial examples being successfully mis-classified by the model. In Fig.~\ref{fig:part4}, we used the attack success rate to evaluate the effect of different attacks. For the defense model, a lower attack success rate means stronger robustness.

We also observed other properties of the network. After SNN is differentiated by Act-BP or Temp-BP, we observed the first-order Taylor expansion of the network with respect to perturbations. The first-order Taylor expansion of SNN can be expressed as~\citep{simon2019first}:
\begin{equation}
    \mathcal{L} (f(\bm{x}+\bm{\delta }),y)-\mathcal{L} (f(\bm{x}),y)=\nabla _{\bm{x}}\mathcal{L} (f(\bm{x}),y) ^ T\bm{\delta }+h\left( \bm{\delta } \right) .
\end{equation}
where $h\left( \bm{\delta } \right)$ is the composition of higher order polynomials. The change in loss before and after perturbation is a scalar, primarily affected by the gradient of clean examples and the perturbation. Therefore, we measured $\lvert \nabla_{\boldsymbol{x}}\mathcal{L} (f(\boldsymbol{x}),y)^T\boldsymbol{\delta } \rvert $. Through a perturbation method, a larger $\lvert \nabla_{\boldsymbol{x}}\mathcal{L} (f(\boldsymbol{x}),y)^T\boldsymbol{\delta } \rvert $ implies a greater change in loss values.

In Fig.~\ref{fig:part3}, we used different encoding and decoding schemes to understand their impact on the robustness of SNNs. We characterized the change in the response of hidden layer neurons due to the presence of perturbations through the spike distance. The distance we use is Victor-Purpura distance, which measures the distance between two spike trains according to the lowest cost of transforming one spike train into another via insertion, deletion, and shifting~\citep{victor1997metric}.

\subsection{Simulation details and parameters}

\noindent \textbf{Details to: Rate-based SNNs gain robustness through synchronization schemes (Fig.~\ref{fig:part1}).} The ReLU-based ANNs were first pretrained for MNIST and Fashion MNIST datasets. The architecture of these models are three-layered network that have 784 input neurons, 100 hidden neurons, and 10 output neurons. The models were converted to SNN as suggested in Eq.~\ref{eq:CVT}. We simulated these SNNs with Poisson coding (Eq.~\ref{eq:poisson_code}), Current coding (Eq.~\ref{eq:current_code}), and Rate-Syn coding (Eq.~\ref{eq:ratesyn_code}) for 30 time steps at the temporal resolution of 0.001 second (1kHz).

In Fig.~\ref{fig:part1}(C), the accuracy of SNN at the $k^{th}$ time step was obtained by averaged the outputs over the time dimension (from the first time step to the $k^{th}$ time step). When we calculated the accuracy with regard to the simulation time, we used the same averaging strategy. 

The attacks towards SNN were performed using ANN with the same parameters as suggested in Sec.~\ref{sec:methods_perturbation}. We performed BIM attacks to converted model with different simulation time with the attack intensity of 0.1, 0.3, 0.5 as suggested in Fig.~\ref{fig:part1}(C). 

In order to discover the intrinsic reasons for robustness, we calculated the distances between the firing rate of hidden neurons in clean and attacked SNN. The firing rate at the $k^{th}$ time step was obtained by recording the spike activity of these neurons from the first time step to the $k^{th}$ time step and calculate the mean. Here two kinds of distances were considered: $l_2$ and $l_\infty$. Besides, we also presented the distances between the ANN activation and SNN mean firing rate under attacks. The ANN activation is also derived from the hidden neurons. Lastly, we compared the instantaneous firing rate of the hidden neuron under attack. This firing rate at the $k^{th}$ time step is obtained by calculating the proportion of firing neurons.

\noindent \textbf{Details to: Dedicated training algorithms help spiking neural networks to further improve robustness (Fig.~\ref{fig:part2}).} We trained SNN using either activation-based back-propagation (Act-BP) and temporal-based back-propagation (Temp-BP) on the Fashion MNIST dataset. These SNNs have three layers, and the hidden neuron number is extended to 500. The results of ANN is from the original ANN model which was later converted to conversion-based SNN following methods mentioned in Eq.~\ref{eq:CVT}. Thus, we have four types of model for the purpose of comparing robust performances. The SNN models for Act-BP and Temp-BP are all trained to receive inputs of totally 10 time steps at 1kHz. In this case, we encode the image input to the corresponding formats. 

The attack we perform in Fig.~\ref{fig:part2} consist of two methods: FGSM (gradient-based attack) and a random deleting perturbation (RandDel). For RandDel, the raw data should be $\{0, 1\}^{T\times C\times H\times W}$ where $T$ is the number of time step, $C, H, W$ is the number of channel, height, and width respectively. The attack strength is determined by a parameter $p$. This parameter is the proportion of randomly deleted spikes. In the actual implementation, we generate a random matrix under a 0-1 uniform distribution. In this matrix, the spikes located on the positions that have random variables smaller than $p$ will be dropped.

% $\lvert \nabla_{\boldsymbol{x}}\mathcal{L} (f(\boldsymbol{x}),y)^T\boldsymbol{\delta } \rvert $
We leveraged first-order Taylor polynomial with respect to pertubations to understand robustness beyond specific attacks. In order to calculate this indicator, the whole calculation process can be divided into three steps. First, through the neural coding method, we got the encoding $I$ of an image $X$. The encoding was later being fed into an SNN to get probabilities for predicting labels. This probability was used to compare with the ground truth label. For the networks trained by Act-BP and Temp-BP, the derivation of gradient w.r.t inputs is the same as the backpropagation method for training them. For the converted SNNs, we applied Act-BP to the backpropagation process of these networks. Back-propagation to these networks produced gradients for the input encoding. Afterwards, we obtained the attacked encoding. We used the attack method to obtain the attacked image $\tilde{X}$. Then we used the neural coding method to construct the attacked encoding, and subtracted the encoding of the original and the attacked image to get $I - \tilde{I}$. Finally, the Hadamard product was performed between $I - \tilde{I}$ and the gradient of the original encoding, and the result matrix was summed. The indicator was obtain after calculating the absolute value of the summed value.

In order to test the transferability of SNN attacks under different training schemes, we trained two groups of SNNs with different random seeds. A group of SNNs perform a white-box attack on themselves. The second set was used to generate adversarial samples for the SNNs of the first set. For SNNs based on conversion, we also applied Act-BP to the backpropagation process of these networks.

\noindent \textbf{Details to: Diverse combinations of encoding and decoding criteria enhance the robustness of SNNs (Fig.~\ref{fig:part3}).}

The model used in the experiments shown in Fig.~\ref{fig:part3} has the same network structure as the model used in Fig.~\ref{fig:part2}. In Fig.~\ref{fig:part3}, the SNNs we trained were all based on Act-BP. The networks here all adopted a combination of neural coding and decoding settings. For the input coding, we chose among Current coding, Rate-Syn coding, TTFS coding as introduced before. For the decoding scheme, we chose between Rate and TTFS decoding. The encoding of TTFS is shown in Eq.~\ref{eq:ratesyn_code}. In order to make TTFS decoding differentiable, after the model output the time of the first spike of each neuron, we calculated the softmax probability of this vector as the prediction of the corresponding classes, and calculated cross entropy loss accordingly with ground truth.

In Fig.~\ref{fig:part3}(E), we showed the spike distances of hidden layer neurons before and after perturbation. The specific process is as follows: First, according to the preset input scheme and the pre-trained network, we obtained the spike sequence of 500 neurons in the hidden layer before and after the perturbation. The number of time steps for these spike trains is 10. We calculated the average distance of spike trains before and after the perturbation for any single neuron as suggested by \cite{victor1997metric}. After getting the mean value, we averaged all samples in the test dataset to get the values shown in Fig.~\ref{fig:part3}(E).

We showed heatmaps of the first-order Taylor polynomial of loss after perturbation. The heatmaps were two-dimensional and involved the generation of two adversarial directions $\vec{\boldsymbol{u}}, \vec{\boldsymbol{v}}$. Since the attacks used in Fig.~\ref{fig:part3} were black-box attacks from the same ANN, we could generate the perturbation direction beforehand. When generating the attack direction, we input the original image with normal distribution noise to the attack model, and performed the same calculation as the FGSM attack. After obtaining the adversarial samples, we subtracted the original samples from the adversarial samples to obtain the perturbation direction. Two different random noises generated two different perturbation directions. In Fig.~\ref{fig:part3}, the linear combinations of these two directions were utilized.

\noindent \textbf{Details to: Application to real-world scenarios (Fig.~\ref{fig:part4}).}

The experiments in Fig.~\ref{fig:part4} simulated more models to resist attacked samples. We also applied more perturbation methods. First, we implemented an attack of Robust Physical Perturbations. We ran their open-source optimization-based attack method and obtained adversarial examples for the images within the class of `stop sign' in the German Traffic Sign Recognition Benchmark dataset.

In subsequent experiments, in addition to FGSM and PGD attack methods, we also adopted MIM, OnePixel, FAB, and DeepFool attack methods. We separately trained a ResNet18 model for attack, and generated corresponding adversarial samples based on this model. Since the model for generating adversarial examples is an ANN, no backpropagation of SNN was involved.

The German Traffic Sign Recognition Benchmark dataset contains 43 classes, with a total of 26640 training samples and 39270 test images. Different image varied in sizes, so during data preprocessing, we resized the image to 32$\times$32, and added random rotation and flip for data enhancement. These data will be transformed into corresponding inputs according to the encoding schemes specified by the model configuration. The simulated number of time steps is 10 at 1kHz.

In terms of network structure, some SNNs had the same network structure as RNNs (LSTM and GRU): they all have 3072 input neurons, three layers of hidden neurons each containing 1000 neurons, and 43 output neurons. The first three layers of neurons of RNN are all recurrent units, except that the output layer takes the output of the last time step of the previous layer and obtains predictions of 43 classes through the linear layer. When the RNN adopts the bidirectional setting, the number of parameters is enlarged to about double, as can be seen in Fig.~\ref{fig:part4}(D).

We showed in Fig.~\ref{fig:part4}(E) the adversarial performance when the fused encoding strategy was adopted and tuned. We used the pre-trained SNN version of VGG9 for the three encodings. Each network produces a representation of $T \times 512$ dimension in the penultimate layer. We averaged the three representations and merged them together. This merged feature was fed into a new linear SNN layer to generate 43 predicted outputs.


\bibliography{myarticle}% common bib file


% \section{Acknowledgments}

% We thank xxx for xxx. We acknowledge xxx for helpful discussions. This work is supported by xxx of the National Natural Science Foundation of China. 

% \section{Author Information}

% \subsection{Authors and Affiliations}

% Department of Electrical and Electronic Engineering, University of Hong Kong, Hong Kong, China
% xxxx

% \subsection{Contributions}

% % Z.W. and S.C.W. conceived the work. Z.W., D.S., S.C.W., Y.L., D.W. and W.Z. contributed to the design and development of the models, software and the hardware experiments. S.C.W., Y.L., C.G., D.W. and W.Z. interpreted, analysed and presented the experimental results. Z.W., D.S., S.C.W. and Y.L. wrote the manuscript. All authors discussed the results and implications and commented on the manuscript at all stages. 
% % All authors contributed to the preparation of the manuscript.

% \section{Competing Interests} The authors declare no competing interests.

% \section{Data and Materials Availability} 

% % The source codes of SpikingJelly can be found at its GitHub
% % page: https://github.com/fangwei123456/spikingjelly. All data are available in the main text or the supplementary materials

% The authors declare that the data supporting this study are available within the paper. Source data are provided with this paper.

\end{document}